\documentclass[sigconf,screen]{acmart}

\usepackage{tabularx, multirow}
\usepackage{array}
\usepackage[most]{tcolorbox}
\usepackage{enumitem}
\usepackage{graphics}

\usepackage{balance}

\AtBeginDocument{%
  }

\usepackage{xcolor}
\usepackage{todonotes}

\newcommand{\heng}[1]{\textcolor{blue}{{\it [Heng says: #1]}}}

\newcommand{\shuwei}[1]{\textcolor{red}{{\it [Shu-wei says: #1]}}}

\definecolor{azure}{rgb}{0.94, 1.0, 1.0}
\definecolor{honeydew}{rgb}{0.94, 1.0, 0.94}
\definecolor{lightgray}{rgb}{0.9, 0.9, 0.9}

\acmBooktitle{Companion Proceedings of the 33rd ACM Symposium on the Foundations of Software Engineering (FSE '25), June 23--27, 2025, Trondheim, Norway}
\begin{document}

%\title{LogLSHD: Fast LSH-based Log Parsing with Grouping and DTW Template Extraction}
\title{LogLSHD: Fast Log Parsing with Locality-Sensitive Hashing and Dynamic Time Warping}

\author{Shu-Wei Huang}
\affiliation{%
  \institution{Polytechnique Montréal}
  \city{Montréal}
  \state{Québec}
  \country{Canada}}
\email{shu-wei.huang@polymtl.ca}
\orcid{0009-0003-8074-6282}

\author{Xingfang Wu}
\affiliation{%
  \institution{Polytechnique Montréal}
  \city{Montréal}
  \state{Québec}
  \country{Canada}
}
\email{xingfang.wu@polymtl.ca}
\orcid{0000-0001-7040-3751}

\author{Heng Li}
\affiliation{%
  \institution{Polytechnique Montréal}
  \city{Montréal}
  \state{Québec}
  \country{Canada}
}
\email{heng.li@polymtl.ca}
\orcid{0000-0001-5441-6763}

\renewcommand{\shortauthors}{Huang et al.}

\begin{abstract}
% Research Background & Motivation
Large-scale software systems generate vast volumes of system logs that are essential for monitoring, diagnosing, and performance optimization. However, the unstructured nature and ever-growing scale of these logs present significant challenges for manual analysis and automated downstream tasks such as anomaly detection. Log parsing addresses these challenges by converting raw logs into structured formats, enabling efficient log analysis. Despite its importance, existing log parsing methods suffer from limitations in efficiency and scalability, due to the large size of log data and their heterogeneous formats. %, the need for manual configuration, and the computational demands of processing massive datasets.
% Research Objectives
To overcome these challenges, %this study aims to develop an efficient log parsing method that overcomes the limitations of current approaches while improving the accuracy and performance of log parsing.
% Research Method
this study proposes a log parsing approach, LogLSHD, which %employs a grouping strategy to effectively reduce data dimensionality. The method 
leverages Locality-Sensitive Hashing (LSH) to group similar logs and integrates Dynamic Time Warping (DTW) to enhance the accuracy of template extraction.
% Results and Contributions
LogLSHD demonstrates exceptional efficiency in parsing time, significantly outperforming state-of-the-art methods. For example, compared to Drain, LogLSHD reduces the average parsing time by 73\% while increasing the average parsing accuracy by 15\%  on the LogHub 2.0 benchmark. %~\xingfang{ on the LogHub 2.0 benchmark?}\shuwei{added} %including AEL and Drain, with reductions of approximately 93\% and 73\% parsing time, respectively. It achieves stable accuracy metrics(e.g., PA, GA) and maintains linear scalability.

\end{abstract}




\maketitle

\section{Introduction}
\label{sec:introductions}
%\textbf{Context.} 

Large-scale software systems are integral to modern computing, encompassing diverse domains such as cloud platforms, Internet of Things (IoT) systems, web applications, telecommunication networks, supercomputers, and more. The scale and complexity of such systems make their monitoring and maintenance increasingly challenging ~\cite{langelier2005visualization}. System logs are essential for capturing detailed runtime information, enabling effective system analysis, maintenance, and optimization. They are crucial for monitoring behaviors~\cite{gadler2017mining, wang2021would, aghili2023studying}, diagnosing issues~\cite{messaoudi2021log, chen2021demystifying, chen2021pathidea, li2021studying}, and improving performance ~\cite{nagaraj2012structured, liao2021locating, li2018adopting}. 

System logs are semi-structured and generated in large volumes, typically consisting of a message header and the log content, as exemplified in Table~\ref{tab:log_example}. 
While the header often contains structured metadata, the log content is usually a natural language record, making it difficult to extract meaningful information. This semistructured nature, combined with the sheer volume of logs, poses significant challenges for manual processing and limits their direct use in various tasks (e.g., anomaly detection). Log parsing addresses these challenges by converting raw log data into structured formats represented by a limited number of log templates and their corresponding parameters (i.e., log variables), enabling automated log analysis ~\cite{he2017towards, zhu2019tools, he2021survey, zhu2023loghub, khan2024impact, wu2023effectiveness}. The structured data produced through log parsing can then be utilized as input for a wide range of downstream log analysis tasks, including those leveraging machine learning (ML) (e.g., \cite{he2016experience}) or deep learning (DL) models (e.g., \cite{le2022log}). %, facilitating a wide range of downstream log analysis tasks.

\begin{table}[!t]
    \caption{An example log message and its components.} % Log parsing aims to extract static log templates and dynamic variables from unstructured log content.}
    \label{tab:log_example}
    %\footnotesize
    \centering
    \resizebox{\columnwidth}{!}{
    \begin{tabular}{@{} p{0.21\columnwidth} p{0.83\columnwidth} @{}}
    \hline
    \textbf{Log message}    
    & 
    2025-01-30 18:01:01 INFO Found block rdd\_42\_20 locally
    %\colorbox{gray!30}{Found block \textit{rdd\_42\_20}} \colorbox{gray!30}{locally}
    %\tablecolorbox[lightgray]{2025-01-30 18:01:01 INFO Found block \textit{rdd\_42\_20} locally} 
    \\ \hline  
    
    \textbf{Header}     
    & 
    2025-01-30 18:01:01 INFO
    %\tablecolorbox[lightgray]{2025-01-30 18:01:01 INFO} 
    \\ \hline 
    
    \textbf{Log content}
    & 
    Found block \colorbox{gray!30}{rdd\_42\_20} locally
   % \tablecolorbox[lightgray]{Found block \textit{rdd\_42\_20} locally}
    \\ \hline 
    %\textbf{Template (static)}
    \textbf{Template}
    & 
    Found block \colorbox{gray!30}{<$*$>} locally
   % \tablecolorbox[lightgray]{Found block \textit{rdd\_42\_20} locally}
    \\ \hline 
    %\textbf{Variable (dynamic)}
    \textbf{Variable}
    & 
    \colorbox{gray!30}{rdd\_42\_20}
   % \tablecolorbox[lightgray]{Found block \textit{rdd\_42\_20} locally}
    \\ \hline 
    \end{tabular}
    }
% }
\vspace{-2mm}
\end{table}

Traditionally, practitioners write \emph{ad hoc} regular expressions to parse logs~\cite{zhang2023system, qin2024preprocessing}. These regular expressions may be designed such that they match the log templates, for example, templates derived from logging statements in the source code~\cite{xu2009detecting,schipper2019tracing}. %\xingfang{I am not sure if it is true. The regular expressions are usually designed to match the variables rather than the log templates.}
However, a typical modern software system can generate logs with a wide variety of templates: for example, the Apache Hadoop system\footnote{https://hadoop.apache.org/} has more than two hundred log templates. The large number of log templates is complicated by the frequent update of logging statements in the source code~\cite{li2017towards}, making it difficult to design and maintain regular expressions for log parsing.
To address the challenge of log parsing, prior work has proposed various log parsers (e.g., Drain~\cite{he2017drain}, AEL~\cite{jiang2008abstracting}, Logram~\cite{dai2020logram}) and benchmarks (e.g., the Loghub~\cite{zhu2023loghub} and Loghub-2.0~\cite{jiang2024large} benchmarks) for automated log parsing. 

The existing log parsers can be broadly divided into two categories: statistic-based approaches and semantic-based approaches~\cite{jiang2024large}. Statistic-based methods rely on predefined statistical patterns (e.g., token frequencies~\cite{dai2020logram}) to extract templates, making them ineffective at handling complex or heterogeneous logs. % from various software systems. 
In contrast, semantic-based approaches (or learning-based, such as LogPPT~\cite{le2022log}) leverage deep neural networks such as large language models to learn the templates from unstructured logs, achieving better parsing accuracy. % semantic information from parameters and templates during parsing to derive structural outputs. 
However, these semantic-based methods typically have higher requirements for computational resources (e.g., GPUs) and take longer time~\cite{jiang2024large}; besides, they typically require labeled data for training~\cite{liu2022uniparser}, fine-tuning~\cite{ma2024llmparser}, or few-shot learning~\cite{le2022log}. %  and introduce additional processing overhead.
% \todo{discuss the pros and cons of these two categories}.
Among the various proposed log parsers, Drain~\cite{he2017drain}, a statistic-based approach leveraging a fixed-depth parse tree, remains the state of the art due to its relatively high parsing accuracy and efficiency~\cite{jiang2024large}. %However, according to a recent benchmark~\cite{jiang2024large}, existing log parsers still fail to achieve a combination of  

%Prior work highlighted several challenges of log parsing~\cite{he2021survey, zhang2023system}. First, the complexity of software systems, resulting in a wide variety of event templates
%Log parsing is essential but encounters several challenges. For example, \citet{he2021survey} identified three key challenges in log parsing, including 1) the substantial volume of logs, which requires significant effort to manually construct regular expressions; 2) the complexity of software systems, resulting in a wide variety of event templates; and 3) the frequent updates in software, necessitating regular modifications to logging statements.
\begin{comment}
\begin{enumerate}
    \item The substantial volume of logs, which requires significant effort to manually construct regular expressions.
    \item The complexity of software systems, resulting in a wide variety of event templates.
    \item The frequent updates in software, necessitating regular modifications to logging statements.
\end{enumerate}
\end{comment}
%\citet{zhang2023system} also highlighted key challenges in log parsing, including the limited research on automatic parameter tuning, which holds potential for adaptive solutions without human intervention. They emphasized that privacy concerns restrict the availability of public log datasets, and the increasing scale of logs imposes significant resource demands on parsers. Additionally, they suggested combining multiple parsers to leverage their strengths and reduce reliance on human input for satisfactory performance.

Building on these insights, this study proposes LogLSHD as an efficient and adaptive log parsing method to address these challenges. LogLSHD leverages Locality-Sensitive Hashing (LSH) and Dynamic Time Warping (DTW) to tackle scalability and accuracy concerns. Specifically, LSH enables fast similarity search by hashing similar items into the same buckets to quickly find similar groups of logs, and DTW finds the optimal alignment within each group to extract log templates. %\xingfang{Specifically, LSH enables fast similarity search by hashing similar items into the same buckets to quickly find similar groups of logs, and DTW finds the optimal alignment within each group to extract log templates.}\shuwei{added}
% Specifically, it employs a grouping strategy to reduce log data dimensionality\xingfang{not clear what it means here.}, utilizes LSH for rapid classification of similar logs, and integrates DTW to enhance the accuracy of template extraction.
Our evaluation results indicate that LogLSHD outperforms the state-of-the-art log parsers: for example, it achieves a better parsing accuracy than Drain and only needs 1/4 of its parsing time on average. This study further explores the performance and applicability of LogLSHD by addressing the following research questions:

%\heng{Shu-Wei's research questions: RQ1) What's the accuracy of the LSH-based parsing approach? Different grouping/parsing accuracy metrics; the impact of different components/techniques (e.g., pre-filtering, etc.). RQ2) What's the efficiency of the LSH-based parsing approach? the impact of different components/techniques (e.g., pre-filtering, etc.) RQ3) How do the different characteristics of log datasets impact the parsing accuracy and efficiency? Number of unique templates, size (number of logs), repetitiveness (entropy, or simply compression ratio)}

\begin{itemize}[leftmargin=*,topsep=0pt]
    % \item \textbf{RQ1: What is the performance of the LSH-based parsing approach?} 
    \item \textbf{RQ1: What is the performance of LogLSHD?} 
    We compare LogLSHD with 13 statistic-based baseline log parsers on the Loghub-2.0\footnote{\url{https://github.com/logpai/loghub-2.0}} benchmark, evaluating both accuracy and efficiency.
    \item \textbf{RQ2: What is the impact of different grouping strategies on the performance of LogLSHD?} 
    We conducted experiments on the Loghub-2.0 benchmark to examine the impact of different grouping strategies on the performance of the LSH-based parsing approach, analyzing how these strategies influence accuracy.
    \item \textbf{RQ3: What is the impact of the DTW-based template extraction method in LogLSHD?} We examine whether DTW could enhance the performance of template extraction processes, aiming to evaluate the effectiveness of DTW in handling diverse and complex log formats. 
\end{itemize}

The main contributions of this work include: 1) a new log parsing algorithm that is several times faster and more accurate than state-of-the-art statistic-based log parsers; 2) a thorough evaluation of the performance of the proposed parser on multiple datasets and different configurations. We share a replication package for future work to replicate or build on our work\footnote{\url{https://github.com/mooselab/LogLSHD}}. %\heng{add the replication package as a footnote or reference}\shuwei{added}.
% This study introduces a log parser combining LSH with grouping strategies and DTW for efficient parsing. Compared to AEL and Drain, the proposed method reduces parsing time by 93\% and 73\%, respectively, while improving template extraction accuracy by 22\% on average. %Our implementations are publicly available as open source\footnote{\url{under maintenance}}.

The remainder of the paper is organized as follows. Section~\ref{sec:motivation} discusses the background and motivation of our work. %, for proposing our method, which addresses several challenges of previous approaches, including efficiency, the high dimensionality of log features, the lack of adaptability, and other key factors related to effective log parsing.
We then describe our methodology in Section~\ref{sec:methodology}, followed by our experiment design in Section~\ref{sec:study_design}. %, covering the details of LogLSHD approach, the datasets, the parsers, the experimental variables, and the evaluation metrics. 
Our research questions are answered in detail in Section~\ref{sec:study_results}. Section~\ref{sec:threats_to_validity} discusses the validity threats, % we may encounter in the research. We then discuss 
followed by a discussion of previous research related to our study in Section~\ref{sec:related_work}. Finally, our conclusions are made in Section~\ref{sec:conclusion}. 

%(i.e., effectiveness and efficiency)
%\section{Motivation}
\section{Background and Motivation}
\label{sec:motivation}
%In this section, we introduce the motivation of our study. 

% \subsection{The importance of efficiency of Log Parsing}
\subsection{Trade-off between efficiency and accuracy}
\label{sec:tradeoff}
% \label{sec:importance of efficiency}

%Current log parsing methods can be broadly categorized into statistic-based and semantic-based approaches. Statistic-based methods, such as AEL~\cite{jiang2008abstracting} and Drain~\cite{he2017drain}, rely on predefined rules or patterns to extract templates. These methods are known for their simplicity and efficiency when applied to structured logs. However, they often struggle to handle heterogeneous log patterns and formats, resulting in low accuracy and reduced effectiveness in log analysis.
%\xingfang{it's the drawback of the semantic-based methods: or scale effectively to large volumes of data. }
%Semantic-based methods, on the other hand, leverage semantics extracted from log messages to identify patterns and relationships in logs~\cite{li2023did}. Examples include UniParser~\cite{liu2022uniparser}, which utilizes self-supervised learning for robust template extraction, and LogPPT~\cite{le2023log}, which employs a pre-training framework to improve parsing accuracy. While these methods offer greater flexibility and adaptability to varying log data, they typically require more complex computations, leading to increased time overhead, which becomes especially prominent in large-scale systems~\cite{du2017deeplog}.
% amounts of annotated data for supervised training, which can be costly and time-consuming to obtain, resulting in slower parsing speeds, especially in large-scale systems~\cite{du2017deeplog}. \xingfang{training do not lead to slower parsing speed.}

As introduced in Section~\ref{sec:introductions}, current log parsing methods can be broadly categorized into statistic-based and semantic-based approaches: the former (e.g., Drain) typically being more efficient but less accurate, while the later (e.g., LogPPT) being more accurate but less efficient~\cite{jiang2024large}. 
Given that modern systems generate overwhelming volumes of logs—often reaching terabytes per day—the efficiency of log parsing has become a critical requirement. Efficient and accurate log parsing not only ensures the timely extraction of structured information but also enables real-time insights and minimizes delays in downstream tasks such as anomaly detection and debugging~\cite{xu2009online, zhu2019tools}. To address the limitations of current methods, this study emphasizes enhancing the efficiency of log parsing while aiming to achieve a good balance between scalability and accuracy. %the accuracy and efficiency of statistic-based approaches, aiming to bridge the gap between scalability and performance.

% \subsection{Potential of LSH}

\subsection{Efficiency issues in log grouping} % caused by high dimensionality}

The log parsing process typically involves several stages\footnote{Exceptions exist, especially semantic-based approaches may predict the log variables directly (e.g., LogPPT)}: %. , as illustrated in Figure \ref{fig:general parsing process}. These stages include 
1) a preprocessing step to clean and tokenize raw logs, 2) log grouping/classification to group similar log entries, 3) postprocessing for grouping refinement, and finally, 4) template extraction, which focuses on identifying structured patterns or templates from unstructured log messages.

\begin{comment}
\begin{figure}
    \centering
    \includegraphics[width=1\linewidth]{figures/overview of the general log parsing process.png}
    \caption{Overview of the general log parsing process.}
    \Description{}
    \label{fig:general parsing process}
    \vspace{-4mm}
\end{figure}
\end{comment}

%\xingfang{Modified. but still needs improvement: need to mention mechanisms of previous approaches.}
Complex software systems may generate large volumes of log messages with numerous and evolving templates. These templates often contain a large number of unique tokens, posing significant challenges for grouping algorithms that rely on token-based features, the dimensionality of which is determined by the unique tokens within log templates. Therefore, efficiently identifying patterns within high-dimensional and large-volume log data is critical.

We consider \textbf{Locality-sensitive hashing} (\textbf{LSH})~\cite{datar2004locality} as a highly efficient approach for grouping log messages into template groups. LSH is a dimensionality reduction technique designed to hash similar data points into the same buckets with high probability while minimizing collisions for dissimilar points. This property makes LSH particularly effective in tasks that require approximate nearest neighbor (ANN) search in high-dimensional spaces (e.g., search for similar logs to form groups). %The primary advantages of LSH include its sub-linear query performance and its ability to provide theoretical guarantees on query accuracy.

%\xingfang{the following is unrelated.}\shuwei{this section can be removed}
%LSH demonstrates extensive applicability across diverse domains. In audio processing, it facilitates efficient music retrieval and anomaly detection. In image and video processing, LSH supports content-based retrieval, video anomaly detection, and encrypted image indexing~\cite{kulis2009kernelized,zhang2016video,xia2016privacy}. The security and privacy domain benefits from LSH in malware clustering, spam detection, and data anonymization~\cite{opricsa2014locality,ozawa2015online,zhang2016scalable}. For data mining, LSH accelerates similarity matrix generation, clustering, and anomaly detection~\cite{koga2007fast,ravichandran2005randomized,zhang2017lshiforest}. Its application in document processing, such as similarity searches and near-duplicate detection~\cite{jiang2011semi,turrado2019locating,li2014large}, offers a foundation for exploring LSH's potential in log parsing. In our proposed approach, we adopt the LSH algorithm with the aim of making the grouping process both efficient and accurate.

% \subsection{Methods of template extraction}
\subsection{Inaccuracy issues in log template extraction}

Many statistic-based log parsers rely on comparing the tokens of log entries to extract log templates~\cite{zhang2023system}. 
%Token comparison is a widely used technique in log parsing due to its simplicity and effectiveness. 
%It enables accurate identification of static and dynamic components by directly comparing tokens in structured or semi-structured log entries. 
For example, Drain~\cite{he2017drain} compares tokens in the same positions within a fixed-depth tree structure to distinguish static and dynamic components and extract log templates. 
%, ensuring efficient classification and template extraction. 
However, token comparison has limitations when handling highly unstructured or noisy logs, where inconsistent token boundaries or significant variations can lead to inaccuracies. Additionally, the reliance on predefined token positions may reduce adaptability to diverse log formats.

To address these challenges, we adopt Dynamic Time Warping (DTW), which extends the capability of token comparison by aligning sequences based on their overall structure rather than fixed positions. DTW identifies the optimal alignment path between log entries, capturing shared patterns while accounting for variations in sequence length and order. This flexibility allows DTW to handle diverse and noisy logs more effectively, making it a robust solution for template extraction in scenarios where token comparison alone may fail. By combining DTW with postprocessing, our approach ensures high accuracy and adaptability in log parsing tasks.

\section{Methodology}
\label{sec:methodology}

\begin{figure}
    \centering
    \includegraphics[width=1\linewidth]{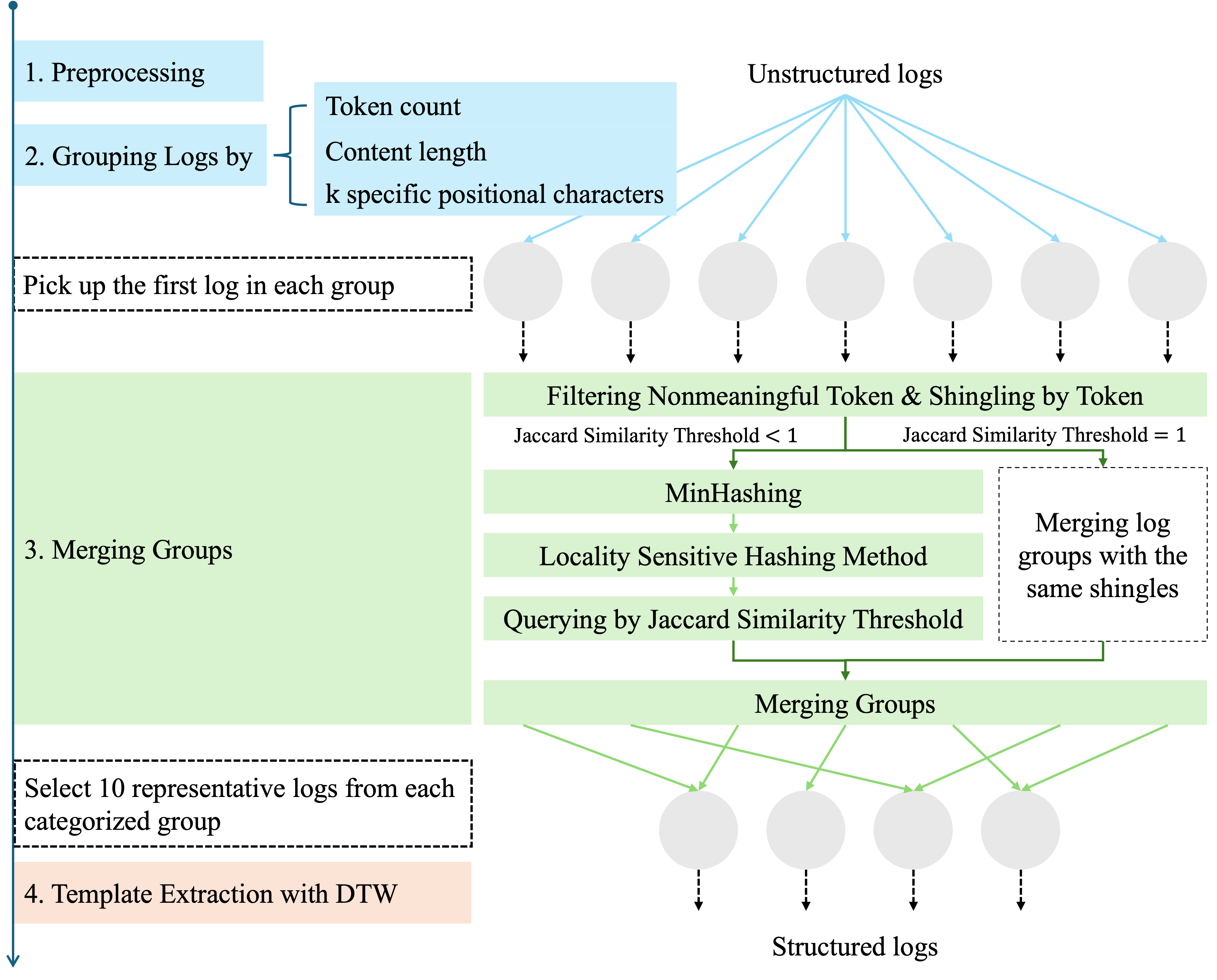}
    \caption{The structure of LogLSHD.}
    \Description{}
    \label{fig:struc_LogLSHD}
    \vspace{-2mm}
\end{figure}

In this section, we present the methodology of our proposed approach, LogLSHD, which aims to efficiently process and group log data for template extraction. As illustrated in Figure~\ref{fig:struc_LogLSHD}, the approach is structured into four main stages: preprocessing, log grouping, group merging, %data classification, 
and template extraction. Each stage is designed to address specific challenges inherent in log data, such as the large log volume, the high-dimensional token space, and the diversity in log templates. %, and the need for efficient grouping. 
The following sections detail the techniques and processes employed in each stage, from log preprocessing and grouping to the final extraction of log templates.

\subsection{\textbf{Preprocessing}}

We followed the same preprocessing step provided by Loghub-2.0~\cite{jiang2024large}. Specifically, raw log messages were preprocessed to replace certain common variables, such as IP addresses and domain names, using prior knowledge based on domains specified by Loghub-2.0. Domain-specific regular expressions were used to match common patterns and replace tokens with the placeholder \textless{}*\textgreater{}. For example, the regex pattern \texttt{(\textbackslash{}d+\textbackslash{}.){3}\textbackslash{}d+} was employed to replace IP addresses in datasets like Hadoop, Spark, Thunderbird, Linux, Apache, and OpenSSH. %These regular expressions are lightweight and focus on token-level matching, ensuring generalization across datasets.

\subsection{\textbf{Initial Log Grouping}} \label{sec:initial-grouping}
%\xingfang{should be checked. it is not clear what is the relationship between the features here and features mentioned in RQ2. Especially, the five-character segmentation mentioned in 5.2.2 is not explained in here, whether it is part of this method? What is the sequence of Token count segmentation, content length segmentation and positional char seg (seem like five-character segmentation belongs to this class?)? Besides, the use ``grouping" instead of ``segment" to avoid inconsistency.}

Before applying LSH-based grouping (see \ref{sec:group-merging}), we apply simple heuristics to initially group unstructured log messages that we expect to belong to the same groups. %are initially grouped based on 
If two log matches match all three pre-defined criteria: token count, content length, and specific positional characters, we assume that they belong to the same group. \textbf{Token count} matches logs with the same number of tokens, assuming that logs with the same token counts may share structural patterns. \textbf{Content length} matches logs based on their overall length (i.e., number of characters). %, grouping messages of similar lengths together. 
\textbf{Specific-position characters} matches logs with the same characters at specific positions (i.e., first, middle or last character). We further discuss this criterion in Section \ref{sec:grouping_strategies}.
%predefined characters, such as commas or colons, appearing at certain positions within the message, further refining the grouping process. In LogLSHD, we sample several characters located at specific positions in the logs as one of the grouping criteria, the details are provided in Section \ref{sec:grouping_strategies}.
% \shuwei{is this make it clearer?}
This initial grouping approach ensures that logs with similar structures are efficiently grouped together, as the three criteria can be implemented rather efficiently (e.g., checking string length or finding the character in a fixed position of a string). %, forming the foundation for subsequent classification. 
%To further streamline the process and 
To prevent redundant processing in later stages, we choose the first log from each initial group as a representative.

\subsection{\textbf{Merging Initial Groups into Log Clusters}} \label{sec:group-merging}
In this step, we leverage a classic LSH-based approach~\cite{leskovec2020mining} to merge the initial log groups, or more precisely, to further cluster the representative logs\footnote{We use ``log grouping'' in the rest of this subsection for simplicity.} from the initial groups obtained from the previous step. % based on their similarity. 
%To achieve this, LSH is applied to the log lines obtained from the preprocessing stage. Before performing LSH, the log lines are transformed into token shingles.
% \xingfang{hard to understand what are the shingles here.}
Before performing LSH, a token filtering mechanism was implemented to ensure that only meaningful tokens were retained; noise and irrelevant information were filtered out. Specifically, a regular expression, \texttt{\^{}[a-zA-Z]+[.,]*\$}, was employed to %match tokens that adhered to a strict pattern. This pattern ensured 
ensure that only tokens consisting entirely of alphabetic characters were preserved, while optionally allowing a single trailing punctuation mark, such as a period (.) or a comma (,). 
% \shuwei{token-level shingle=split by token}\xingfang{seems like the shingling process is just a filtering process to filter out tokens that are not words or words plus punctuation. not sure if I understand correctly.}

The LSH-based approach consists of four steps: shingling, MinHashing, LSH hashing, and similarity-based grouping~\cite{leskovec2020mining}. We leveraged a public implementation of LSH~\cite{minhashlsh} in our approach. 

% The LSH approach we’re exploring consists of a three-step process. First, we convert text to sparse vectors using k-shingling (and one-hot encoding), then use minhashing to create ‘signatures’ — which are passed onto our LSH process to weed out candidate pairs.

\noindent \textbf{Shingling}. 
The Shingling step converts a string of text into a set of ``shingles'': a $k$-shingle is a $k$-gram substring~\cite{leskovec2020mining}. In our case, we use a token-based shingling: each log message is converted into a set of tokens; each shingle is a token. Then, a sparse vector is used to represent the set of shingles (tokens) for each log message: the length of the vector is the vocabulary size of all shingle sets (i.e., obtained from all log messages); to represent a log message, for its set of shingles, we set the corresponding values of the vector as 1 and other values as 0 (similar to one-hot encoding).

%During filtering, a token-level shingle was simultaneously applied, generating token-based segments from the logs. 

\noindent \textbf{MinHashing}. 
The MinHashing step converts the sparse vectors obtained from the Shingling step into dense vectors of fixed length: signatures~\cite{leskovec2020mining}. It would be much more efficient to compare the dense vectors than the sparse vectors.
Specifically, considering a dense vector of length $d$, each value of the vector is obtained through using a different \emph{minhashing} function~\cite{leskovec2020mining} to convert (hash) the sparse vector into an integer: there are $d$ different \emph{minhashing} functions in total. The signature length $d$ is a configurable parameter; we fix it at 50 as 50 integers can produce sufficient combinations to distinguish different log templates.
%We use the default $d$ of D provided by our used LSH implementation~\cite{minhashlsh}.
%The token shingles are processed through a MinHashing step, where each tokenized log is represented as a signature of fixed length. 
%MinHashing is a technique used to efficiently estimate the similarity between sets by generating compact signatures through a series of hash functions, which reduce the dimensionality of the data and facilitate faster comparison. We adopt MinHashing here to capture the patterns of the log while significantly reducing the dimensionality of the log features. 

\noindent \textbf{LSH hashing}.
LSH hashing aims to hash similar signatures obtained in the last step into the same bucket (in other words, causing hashing collisions, in a favorable way)~\cite{leskovec2020mining}. 
However, it is difficult to find hashing functions that perfectly hash similar (but different) signatures into the same bucket.
Thus, we use banded LSH hashing~\cite{leskovec2020mining}: each signature is divided to $b$ equal-sized bands, and a hash function is then used to hash each band into a bucket. If any pair of bands from two log messages are mapped into the same bucket (i.e., a collision), we consider the entire two log messages as \emph{candidate pairs} to be merged into the same cluster. $b$ is a configurable parameter; we use the parameter optimization provided by our used LSH implementation~\cite{minhashlsh}. Although this any-part-matching method can cause many false positives, this process helps identify a relatively small number of log message pairs (i.e., \emph{candidate pairs}) that are subject to further check for their actual similarity. 
%The MinHash signatures are used to estimate the Jaccard similarity between the logs, and based on these signatures, LSH efficiently groups similar logs by hashing them into buckets.
% \xingfang{the logical relationships among the techniques are missing. LSH actually use MinHashing, and features from MinHashing are used to estimate/approximate Jaccard similarity. All the places mentioning the Jaccard similarity should be checked: whether they are actually Jaccard similarity or the similarity calculated with signature of MinHash.}\shuwei{Jaccard similarity is calculated by MinHash signatures in this paper.}

% \xingfang{the problem here is that the wording confuses the threshold and similarity value. Threshold is 1 here. T\_jaccard is not actually threshold: it is the similarity value. Besides, the Jaccard similarity and the similarity of MinHash signatures are also confused.}
\noindent \textbf{Similarity-based grouping}.
The \emph{candidate pairs} obtained in the last step then go through to a more rigorous comparison using an empirically determined Jaccard similarity threshold, $T_{\text{Jaccard}}$. If the Jaccard similarity between the signatures of a \emph{candidate pair} %, as defined in Equation~\ref{eq:jaccard}, 
exceeds this threshold, they will be identified as similar. 
$T_{\text{Jaccard}}$ is a configurable parameter of our approach.
%This comparison ensures that only logs with sufficient similarity are identified. 
Finally, log messages corresponding to the identified similar pairs are grouped into the same cluster which will then be used for template extraction: a cluster corresponds to a template. %together, merging them into cohesive clusters that reflect shared structural or word patterns.
% \xingfang{not sure if the semantic here is proper.}\shuwei{True! Word is more precise.}

\begin{comment}
\begin{equation}
\label{eq:jaccard}
J(A, B) = \frac{|A \cap B|}{|A \cup B|}
\end{equation}
\end{comment}

For some datasets, after filtering nonmeaningful tokens, most of the variables may be removed, leaving only the static parts of the logs. In this case, exact matching (i.e., $T_{\text{Jaccard}} = 1$) may perform better than similarity-based matching.
%a predefined Jaccard similarity threshold ($T_{\text{Jaccard}}$)\xingfang{Not threshold...I think here is the similarity of the groups $S_{\text{MinHash}}$}. 
Only when $T_{\text{Jaccard}} = 1$, shingles with identical content are directly merged, bypassing LSH for similarity detection. %When $T_{\text{Jaccard}} < 1$, LSH is employed to efficiently group similar shingles into candidate pairs. 
%This approach ensures that exact matches are handled with minimal computational overhead while enabling approximate similarity detection when needed. %, making the process adaptable to specific application requirements.

\begin{comment}
The strategy is formalized as follows: 

\begin{equation}
    \begin{cases} 
    \text{Merge groups by shingle directly}, & \text{if } T_{\text{Jaccard}} = 1, \\
    \text{Merge groups with LSH}, & \text{if } T_{\text{Jaccard}} < 1
    \end{cases}
\end{equation}
\end{comment}

% \xingfang{1 is the threshold. the $T_{\text{Jaccard}}$ is not threshold? It is a similarity.}\shuwei{$T_{\text{Jaccard}}$ is defined as the `Jaccard similarity threshold`. In other words, if the Jaccard similarity between the two groups exceeds this threshold, they will be merged. For example, if $T_{\text{Jaccard}}$=0.8, groups with a Jaccard similarity of 0.8 or higher will be merged.}

\subsection{\textbf{Template Extraction}} 

\begin{figure}
    \centering
    \includegraphics[width=1\linewidth]{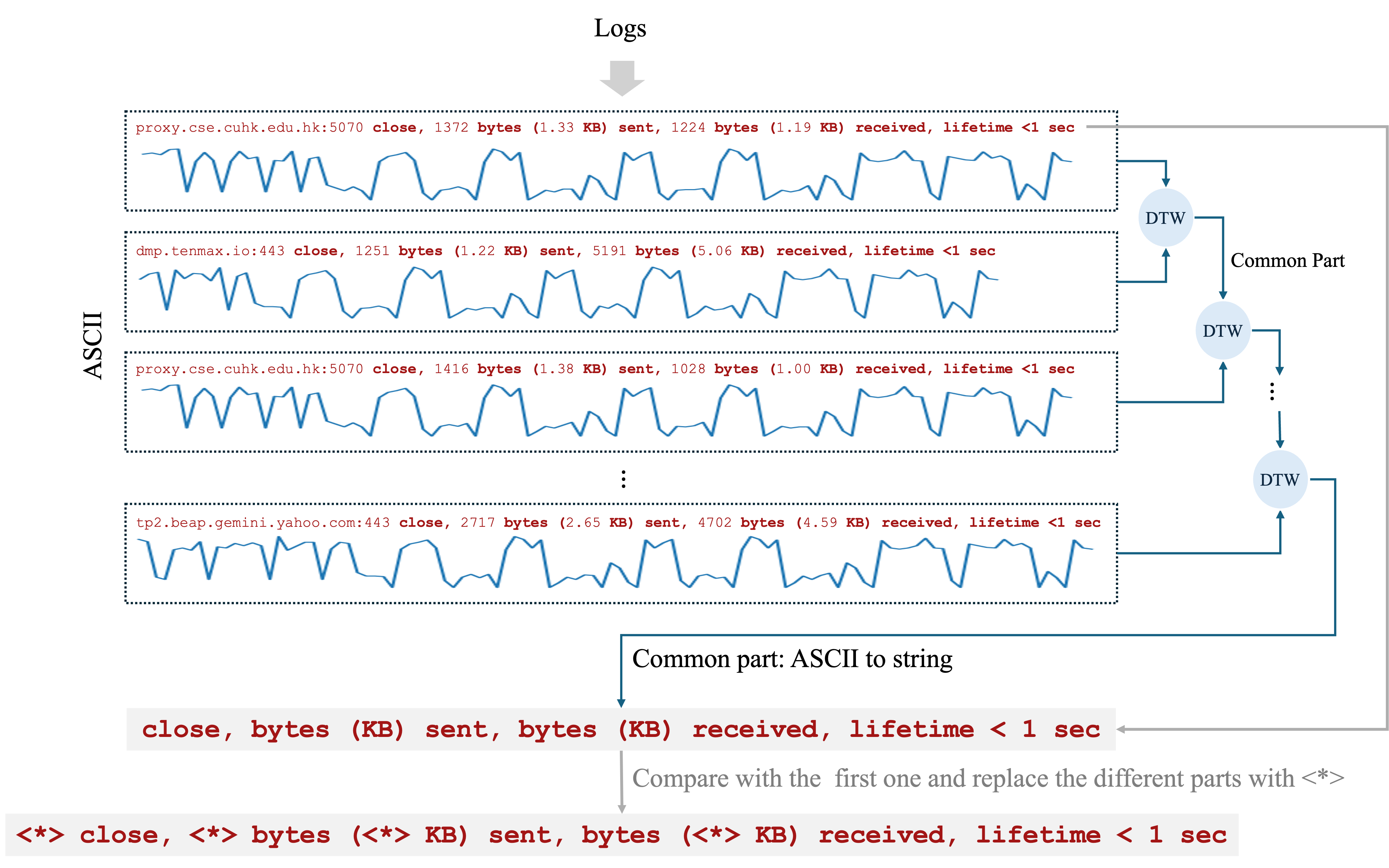}
    \caption{Illustration of template extraction with DTW.}
    \label{fig:DTW}
\end{figure}

After clustering similar logs in the previous step, 10 representative logs are randomly selected from each group for template extraction. We then use the DTW method to identify the longest common subsequence among these logs (see Figure~\ref{fig:DTW}). Each log is converted into an ASCII sequence to facilitate numerical alignment. DTW computes the optimal alignment path through a cost matrix, recording indices of matching characters across sequences. By iteratively applying DTW, the method isolates the static components shared by all logs, while dynamic parts are replaced with placeholders \textless{}*\textgreater{} to form the initial template. Finally, placeholders adjacent to each other %static characters 
are merged together. % while ensuring separation from special tokens (e.g., `:', `=', `[', `]' `(', `)', `\{', and `\}') to maintain the structural integrity of the template.
\section{Experiment Design}
\label{sec:study_design}

In this section, we describe our experiment design, including the experiment environment, evaluation datasets, baselines, experimental process, and evaluation metrics. 
%we first introduce the experimental environment in Section~\ref{sec:experiment_environment}. Details of the datasets and baseline parsers used in our experiments are provided in Section~\ref{sec:datasets} and Section~\ref{sec:baseline_parsers}, respectively. The experimental process is described in Section~\ref{sec:experiment_variables}, followed by an introduction to the evaluation metrics in Section~\ref{sec:eval_metrics}.

\subsection{Experiment Environments}
\label{sec:experiment_environment}

% Our baseline log parser implementation and the log datasets used in this study are sourced from Loghub 2.0\cite{jiang2024large}. 

All the experiments in this paper were conducted on a Colab environment fixed with an AMD EPYC 7B12 processor (8 logical CPUs, 4 cores, 2 threads per core) and 50 GB of RAM, without GPU usage.

\subsection{Datasets}
\label{sec:datasets}

\begin{table}[!t]
    \centering
    \caption{Statistics of Datasets from Loghub-2.0~\cite{jiang2024large}}
    \label{tab:datasets_stats}
    \resizebox{\columnwidth}{!}{%
    \begin{tabular}{m{1.8cm}m{1.5cm}m{1.5cm}m{1.5cm}m{1.2cm}}
    \bottomrule
    \textbf{System Type} & \textbf{Dataset} & \textbf{\#Templates} & \textbf{\begin{tabular}{@{}c@{}}\#Annotated\\Logs\end{tabular}} & \textbf{\begin{tabular}{@{}c@{}}Template\\Density\end{tabular}} \\ \bottomrule
    \multirow{5}{*}{\begin{tabular}{@{}c@{}}Distributed\\systems\end{tabular}} 
          & Hadoop     & 236 & 179,993 & 1.311      \\
          & HDFS       & 46  & 11,167,740 & 0.004   \\
          & OpenStack  & 48  & 207,632 & 0.231     \\
          & Spark      & 236 & 16,075,117 & 0.015  \\
          & Zookeeper  & 89  & 74,273 & 1.198        \\ \hline
    \multirow{3}{*}{\begin{tabular}{@{}c@{}}Supercomputer\\systems\end{tabular}} 
          & BGL         & 320 & 4,631,261 & 0.069      \\
          & HPC         & 74 & 429,987 & 0.172          \\
          & Thunderbird & 1,241 & 16,601,745 & 0.075   \\ \hline
    \multirow{2}{*}{\begin{tabular}{@{}c@{}}Operating\\systems\end{tabular} } 
          & Linux & 338 & 23,921 & 14.130     \\
          & Mac   & 626 & 100,314 & 6.240  \\ \hline
    \multirow{2}{*}{\begin{tabular}{@{}c@{}}Server\\application\end{tabular}} 
          & Apache  & 29 & 51,977 & 0.058   \\
          & OpenSSH & 38 & 638,946 & 0.059  \\ \hline
    \multirow{2}{*}{\begin{tabular}{@{}c@{}}Standalone\\software\end{tabular}} 
          & HealthApp & 156 & 212,394 & 0.353  \\
          & Proxifier & 11 & 21,320 & 0.516   \\ 
    \bottomrule
    \textbf{Average}  &  & \textbf{249.1} & \textbf{3,601,187} & \textbf{1.781} \\ 
    \bottomrule
    \end{tabular}
    }
\end{table}

This study utilizes datasets from Loghub-2.0~\cite{jiang2024large}, an enhanced log parser benchmark collection derived from Loghub-2k~\cite{zhu2023loghub}. The details of the Loghub-2.0 datasets are summarized in Table~\ref{tab:datasets_stats}. Loghub-2.0 consists of 14 datasets spanning diverse software systems, including distributed systems, super-computers, operating systems, server applications, and standalone software. Each dataset contains an average of 3.6 million log entries—a nearly 1,900-fold increase in scale compared to Loghub-2k. The datasets exhibit notable diversity in scale and structure, with annotated logs ranging from 21,320 (Proxifier) to 16,601,745 (Thunderbird) and unique templates spanning from 11 (Proxifier) to 1,241 (Thunderbird). %Log message lengths vary from 3 to 1,264 characters, while the number of variables per log ranges from 0 to 27. 
Template Density represents the number of unique templates per 1,000 annotated logs, as shown in Equation \ref{eq:template＿density} indicating log diversity. A higher density signifies greater template diversity in the dataset. Linux exhibits the highest template density.

\begin{equation}
\label{eq:template＿density}
\text{Template Density} = \frac{\# \text{Templates}}{\# \text{Annotated Logs}} \times 1000
\end{equation}

By incorporating a broader range of log template frequencies and parameter counts, along with large-scale data and diverse system types, Loghub-2.0 provides a realistic and comprehensive representation of modern log data. Its detailed template annotations and high parameter variability capture the diversity and complexity of real-world system logs, enabling rigorous evaluation of parsing methods' scalability, accuracy, and robustness across a wide range of practical scenarios.

\subsection{Baseline Log Parsers}
\label{sec:baseline_parsers}

As introduced in Section \ref{sec:tradeoff}, this study benchmarks its proposed methods against a diverse set of established log parsers to ensure a comprehensive evaluation. The baseline parsers include AEL\cite{jiang2008abstracting}, Drain~\cite{he2017drain}, IPLoM~\cite{makanju2009clustering}, SLCT~\cite{vaarandi2003data}, LogCluster~\cite{vaarandi2015logcluster}, LenMa~\cite{shima2016length}, LogMine~\cite{hamooni2016logmine}, Logram~\cite{dai2020logram}, LogSig~\cite{tang2011logsig}, Spell~\cite{du2016spell}, MoLFI~\cite{messaoudi2018search}, and SHISO~\cite{mizutani2013incremental}. These parsers are representative of statistic-based approaches, which rely on predefined rules or patterns for template extraction. This choice aligns with the focus of the study on improving the accuracy and efficiency of statistic-based methods, making it possible to evaluate their performance comprehensively across diverse datasets.

\subsection{\textbf{Experimental Variables}}
\label{sec:experiment_variables}

\subsubsection{\textbf{Grouping Strategies}}
\label{sec:grouping_strategies}

%\xingfang{not clear whether the five-character segment is part of the method. if so, it should be moved to the methodology. it seems like the five-character segmentation is an extra procedure after the segmentation based on token count and content length.}
Our approach uses three intuitive criteria to perform an initial grouping of log messages: \textit{token count}, \textit{content length}, and \textit{specific-position characters} (see Section~\ref{sec:initial-grouping}).
To explore the influence of different grouping conditions, \textit{token count} and \textit{content length} were employed as the base criteria for the initial log grouping. We then examine how adding the \textit{specific-position characters} criterion impact the parsing performance. Specifically, we consider different combinations of specific-position character matching between two log messages: the first character, the 25\% position, the 50\% position, the 75\% position, and the last character. 
%Based on these criteria, logs were further divided into five-character segments by accumulating specific character positions from each log: the first character, the 25\% position, the 50\% position, the 75\% position, and the last character. This strategy enables the examination of how character-based segmentation affects the accuracy and efficiency of the log grouping process.

\subsubsection{\textbf{Hyperparameters of LSH}}
\label{sec:Hyperparameters of LSH}

In this study, we identify the Jaccard threshold as the primary hyperparameter influencing the overall performance of LogLSHD. To examine the effect of varying similarity degrees, we systematically selected a range of Jaccard thresholds: 1, 0.95, 0.9, 0.85, 0.8, 0.75, 0.7, 0.65, 0.6, 0.55, and 0.5. Additionally, another key hyperparameter, the signature length, was fixed at 50 to balance computational efficiency and grouping accuracy. %By focusing on these hyperparameters, our study provides a comprehensive evaluation of their impact on the log grouping performance and the overall experimental results.

\subsubsection{\textbf{Methods of Template Extraction}}

Instead of using the basic token comparison method to extract templates from grouped log data, this study employs the Dynamic Time Warping (DTW) method to enhance the results, as DTW is proven effective by aligning sequences based on their overall structure rather than fixed positions. Using identical hyperparameters for LSH and a consistent grouping strategy, two approaches—with and without DTW—are compared to evaluate their respective performance.

\subsection{Evaluation Metrics}
\label{sec:eval_metrics}

Assessing the quality of log parsing requires robust metrics, with Grouping Accuracy (GA) and Parsing Accuracy (PA) being the most widely used. GA examines whether log messages with the same ground truth template are grouped together by the parser, while PA focuses on the precision of template extraction by comparing parsed templates with their ground truth counterparts~\cite{zhu2019tools, dai2020logram}. To achieve a more comprehensive evaluation, this study also considers FGA, FTA, and parsing time, following recommendations from prior work~\cite{khan2022guidelines, jiang2024large}.

\noindent\textbf{Grouping Accuracy (GA):}
GA measures the percentage of log messages that are correctly grouped based on their templates. A grouping is considered correct if all log messages sharing the same template in the ground truth are placed in the same group by the parser~\cite{zhu2019tools}. This metric evaluates the log parser's ability to cluster log messages accurately and is formally defined as:
\begin{equation}
GA = \frac{\text{\#Number of correctly grouped logs}}{\text{\#Total number of logs}}
\end{equation}

\noindent\textbf{Parsing Accuracy (PA): }
PA assesses the proportion of templates correctly extracted by the parser~\cite{dai2020logram}. A parsed template is deemed correct if it accurately masks all variable components while preserving the static elements. This metric directly measures the parser's ability to distinguish between the static and dynamic parts of log messages and is defined as:
\begin{equation}
PA = \frac{\text{\#Number of correctly parsed templates}}{\text{\#Total number of logs}}
\end{equation}

\noindent\textbf{{F1-score of Grouping Accuracy (FGA)}:}
Building upon the work of GA, FGA enhances the grouping accuracy by focusing on template-level grouping. It is calculated as the harmonic mean of Precision of Group Accuracy (PGA) and Recall of Group Accuracy (RGA)~\cite{jiang2024large}, which are defined as:

\begin{equation}
PGA=\frac{\text{\#Number of correctly grouped templates}}{\text{\#Number of predicted groups}}
\end{equation}

\begin{equation}
RGA=\frac{\text{\#Number of correctly grouped templates}}{\text{\#Number of ground truth groups}}
\end{equation}

\begin{equation}
FGA=\frac{2\times (PGA\times RGA)}{PGA+RGA}
\end{equation}

By assigning greater weight to uncommon templates, such as error logs, FGA offers a more fair and reliable evaluation of grouping performance compared to GA.

\noindent\textbf{F1-score of Template Accuracy (FTA): }The objective of FTA is to assess the log parser’s ability to accurately identify log templates. A template is deemed correctly identified if it satisfies two conditions: 1) every token within the template aligns with the ground truth, and 2) the associated log messages are accurately assigned ~\cite{khan2022guidelines}. Similar to FGA, FTA is calculated as the harmonic mean of the Precision of Template Accuracy (PTA) and the Recall of Template Accuracy (RTA), defined as:

\begin{equation}
PTA=\frac{\text{\#Number of correctly identified templates}}{\text{\#Number of 
 identified templates}}
\end{equation}

\begin{equation}
RTA=\frac{\text{\#Number of correctly identified templates}}{\text{\#Number of ground truth templates}}
\end{equation}

\begin{equation}
FTA=\frac{2\times (PTA\times RTA)}{PTA+RTA}
\end{equation}

\noindent\textbf{Parsing Time: }
All parsing time measurements were carried out in a fixed environment, as detailed in Section~\ref{sec:experiment_environment}. Parsing Time measures the efficiency of a log parser by quantifying the total time taken to process a log dataset, including reading raw log files and generating the corresponding structured log output.

\section{Experiment Results}
\label{sec:study_results}
In this section, we report our results obtained through the experimental design described in Section~\ref{sec:study_design} and answer the research questions introduced in Section~\ref{sec:introductions}.

\subsection{RQ1: What is the performance of LogLSHD?}
\label{sec:rq1}

\subsubsection{Motivation}
% \noindent \textbf{Motivation} 
The performance of log parsing approaches is crucial for efficient and accurate analysis of large volumes of log data. Previous log parsers have shown limitations in completing tasks within reasonable timeframes, motivating the need for a more efficient approach while maintaining high parsing accuracy. This research question aims to assess the performance of our proposed log parser, LogLSHD, in terms of both accuracy and efficiency.

\subsubsection{Approach}
This research question involves comparing LogLSHD with several baseline log parsers across different datasets.
We use the default parameters of each log parser; for LogLSHD, the default $T_{\text{Jaccard}}$ values (shown in Table~\ref{tab:LogLSHD_acc}) are obtained empirically: they are determined based on the combined Parsing Accuracy (PA) and Grouping Accuracy (GA) ($PA + GA$) for each dataset.
Key evaluation criteria include parsing accuracy and time efficiency. Parsing accuracy is assessed using multiple metrics to evaluate the performance of each approach. %, with a focus on robustness across different datasets. 
Time efficiency is analyzed by comparing parsing times across datasets of varying sizes. Additionally, the experiment explores the impact of the similarity threshold on both parsing accuracy and efficiency, examining how different thresholds affect log grouping and parsing performance.

\subsubsection{Result}
\paragraph{\textbf{Comparison of Parsing Accuracy Metrics}}

\begin{figure}[ht]
    \centering
    \includegraphics[width=1\linewidth]{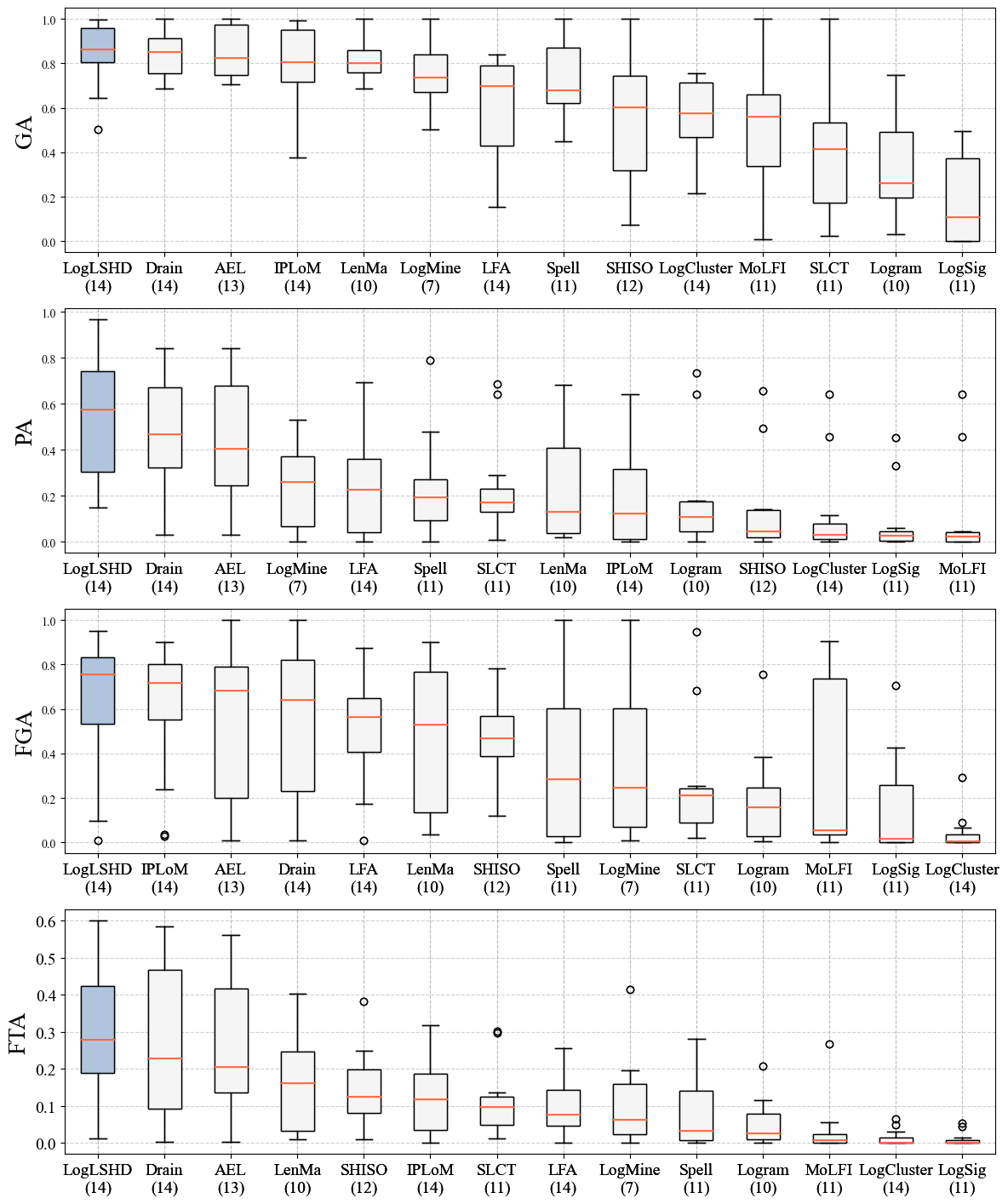}
    \caption{Comparison of log parsing methods with accuracy metrics.}
    \Description{}
    \label{fig:parser comparison with acc metrics}
\end{figure}

According to the experimental results reported by Jiang et al.~\cite{jiang2024large}, not all log parsers could complete their tasks within 12 hours under their experimental settings. Therefore, this study adopts their approach and denotes the number of datasets each log parser can process within 12 hours in parentheses under the parser’s name in Figures~\ref{fig:parser comparison with acc metrics} and~\ref{fig:parser comparison with time}.

As shown in Figure ~\ref{fig:parser comparison with acc metrics} where the accuracy metric distribution of all log parsers are ranked by the median in ascending order from left to right, LogLSHD consistently achieves the best performance across all accuracy metrics. Furthermore, compared to most baselines, LogLSHD exhibits narrower upper and lower bound differences, demonstrating its superior robustness.

\begin{table}[!t]
    \centering
    \caption{Average Performance Across Methods}
    \label{tab:average_metrics}
    \begin{tabular}{lccccc}
        \toprule
        \textbf{Method} & \textbf{PA} & \textbf{GA} & \textbf{FGA} & \textbf{FTA} & \textbf{Parse Time (s)} \\
        \midrule
        LogLSHD & \textbf{0.54} & \textbf{0.85} & \textbf{0.64} & \textbf{0.30} & \textbf{109} \\
        Drain & 0.47 & 0.84 & 0.54 & 0.28 & 405 \\
        AEL & 0.44 & 0.86 & 0.56 & 0.25 & 1654 \\
        % IPLoM & 0.19 & 0.79 & 0.61 & 0.12 & 368 \\
        % LFA & 0.25 & 0.60 & 0.52 & 0.10 & 324 \\
        % SLCT & 0.24 & 0.40 & 0.27 & 0.11 & 2578 \\
        % LenMa & 0.24 & 0.82 & 0.47 & 0.16 & 1511 \\
        % LogMine & 0.24 & 0.75 & 0.37 & 0.12 & 4319 \\
        % Spell & 0.24 & 0.73 & 0.35 & 0.09 & 1228 \\
        % Logram & 0.20 & 0.34 & 0.20 & 0.05 & 1054 \\
        % SHISO & 0.14 & 0.54 & 0.48 & 0.15 & 1413 \\
        % MoLFI & 0.11 & 0.53 & 0.36 & 0.04 & 3718 \\
        % LogCluster & 0.11 & 0.57 & 0.04 & 0.01 & 191 \\
        % LogSig & 0.09 & 0.18 & 0.16 & 0.01 & 2172 \\
        \bottomrule
    \end{tabular}
\end{table}

\begin{table}
    \centering
    \caption{Performance of LogLSHD in four different accuracy metrics.}
    \label{tab:LogLSHD_acc}
    \begin{tabular}{lccccc}
    \bottomrule
    \textbf{Dataset} & \textbf{$T_{\text{Jaccard}}$} & \textbf{GA} & \textbf{PA} & \textbf{FGA} & \textbf{FTA} \\ \bottomrule
    Proxifier & 1.00 & 0.504 & 0.756 & 0.007 & 0.011 \\ \hline
    Linux & 0.65 & 0.791 & 0.514 & 0.656 & 0.187 \\ \hline
    Apache & 0.65 & 0.997 & 0.727 & 0.949 & 0.508 \\ \hline
    Zookeeper & 0.80 & 0.984 & 0.811 & 0.810 & 0.601 \\ \hline
    Hadoop & 0.85 & 0.932 & 0.451 & 0.815 & 0.377 \\ \hline
    HealthApp & 0.65 & 0.852 & 0.745 & 0.513 & 0.258 \\ \hline
    OpenStack & 0.70 & 0.960 & 0.151 & 0.936 & 0.298 \\ \hline
    HPC & 0.70 & 0.859 & 0.682 & 0.841 & 0.524 \\ \hline
    Mac & 0.95 & 0.869 & 0.277 & 0.725 & 0.199 \\ \hline
    OpenSSH & 0.85 & 0.643 & 0.293 & 0.098 & 0.056 \\ \hline
    Spark & 0.90 & 0.859 & 0.637 & 0.835 & 0.438 \\ \hline
    Thunderbird & 0.60 & 0.698 & 0.230 & 0.597 & 0.175 \\ \hline
    BGL & 0.90 & 0.965 & 0.338 & 0.782 & 0.219 \\ \hline
    HDFS & 0.80 & 0.949 & 0.967 & 0.436 & 0.319 \\ \bottomrule
    Average & & 0.847 & 0.541 & 0.643 & 0.298 \\ \bottomrule
    \end{tabular}
\end{table}

\subsubsection{\textbf{Comparison of Parsing Time}}
\begin{figure}[ht]
    \centering
    \includegraphics[width=1\linewidth]{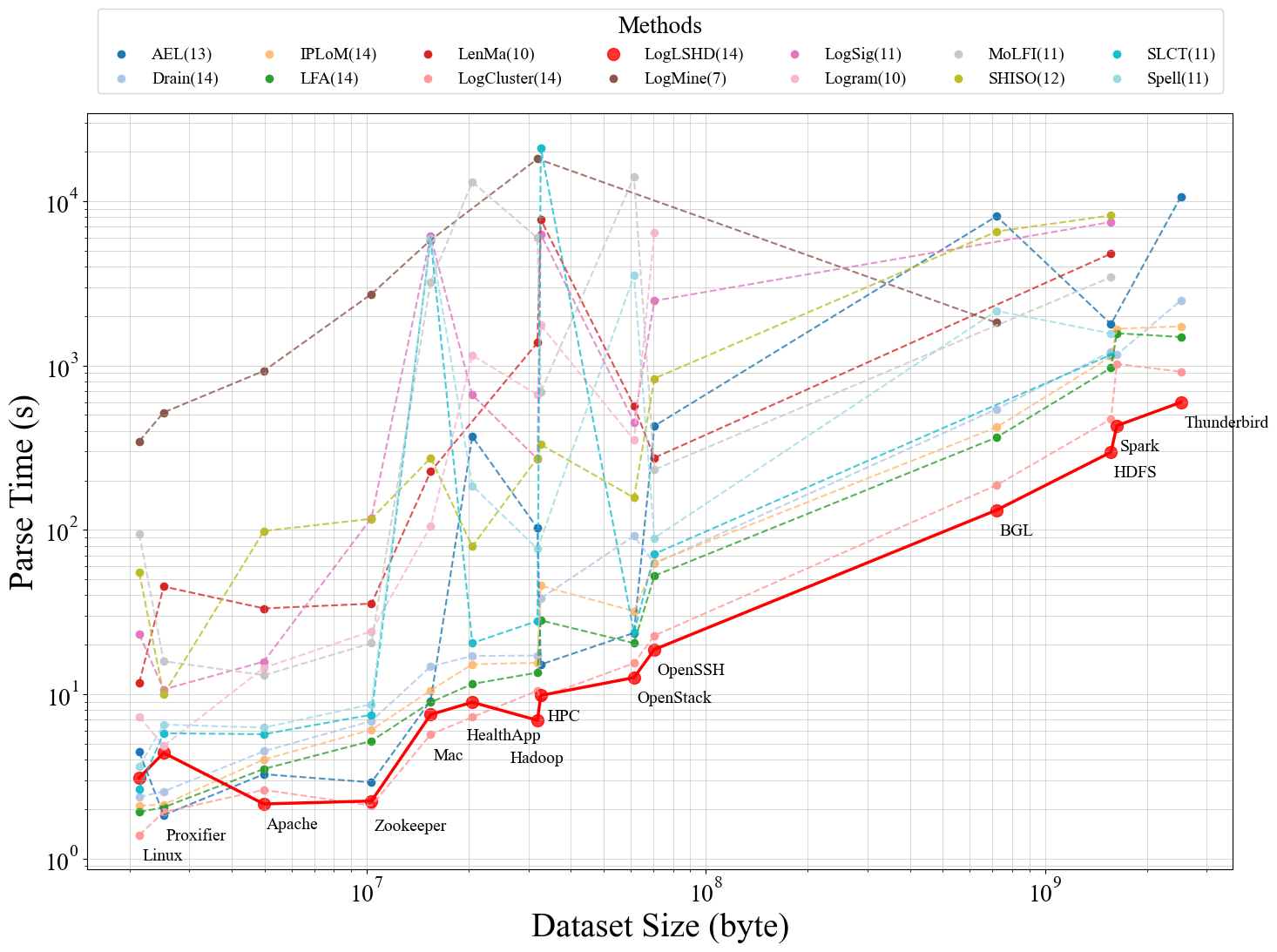}
    \caption{Comparison of log parsing methods with parsing time.}
    \Description{}
    \label{fig:parser comparison with time}
\end{figure}

Figure~\ref{fig:parser comparison with time} illustrates the parsing time of various log parsing methods across different dataset sizes. Each dashed line represents the performance of a specific method, while the red solid line highlights the performance of the proposed LogLSHD method. As dataset sizes increase, all methods show a growth in parsing time, with significant differences in the rate of increase and overall performance. According to Table~\ref{tab:average_metrics}, LogLSHD achieves an average parsing time of only 109 seconds, representing a 93.4\% and 72.8\% improvement in efficiency compared to AEL and Drain, respectively. Notably, when the dataset size grows from $10^7$ to $10^9$, the required parsing time for LogLSHD increases by a factor of $10^2$, demonstrating its consistent and stable scalability with linear growth in parsing time.

\paragraph{\textbf{Challenges with Linux and Proxifier Datasets}}
Interestingly, compared with some other log parsers (e.g., IPLoM or Drain), LogLSHD demonstrates relatively lower efficiency on the Proxifier and Linux datasets. %\xingfang{from the result, it seems like the parsing time for these two datasets is lower than that of other datasets. why does LogLSHD demonstrate relatively lower efficiency? compared with other parsers?}
As shown in Table~\ref{tab:datasets_stats},
% \xingfang{\ref{tab:datasets_stats}?}\shuwei{revised}
Linux has the highest template density. Template extraction using DTW involves independent computations for each template, and a higher template density leads to a significant increase in parsing time.
% \xingfang{I don't understand this from the results.}\shuwei{revised the explanation with Linux result, table 2, section 4.2} 
On the other hand, LogLSHD’s suboptimal performance on Proxifier can be attributed to the presence of two highly similar templates: ``\url{proxy.cse.cuhk.edu.hk:5070} close, 403 bytes sent, 426 bytes received, lifetime <1 sec'' and ``\url{proxy.cse.cuhk.edu.hk:5070} close, 1124 bytes (1.09 KB) sent, 529 bytes received, lifetime <1 sec''. As discussed earlier, the proposed filtering method generates identical shingles for these templates (``close, bytes sent, bytes received, lifetime sec''). This results in the failure of $T_{\text{Jaccard}}<1$ to distinguish them effectively. Although a strict matching criterion ($T_{\text{Jaccard}}=1$) improves accuracy, it also incurs higher computational costs, since directly merging logs with the same shingles can be slower than LSH-based merging as the later reduces the pair-wise comparisons to the small set of \textit{candidate pairs} (see Section~\ref{sec:group-merging}). %, as highlighted in Figure~\ref{fig:parser comparison with time}. 
%This trade-off between accuracy and efficiency underlines the need for careful parameter tuning in practical applications.

\paragraph{\textbf{Impact of Jaccard Threshold on Parsing Performance}}
\begin{figure*}
    \centering
    \includegraphics[width=1\linewidth]{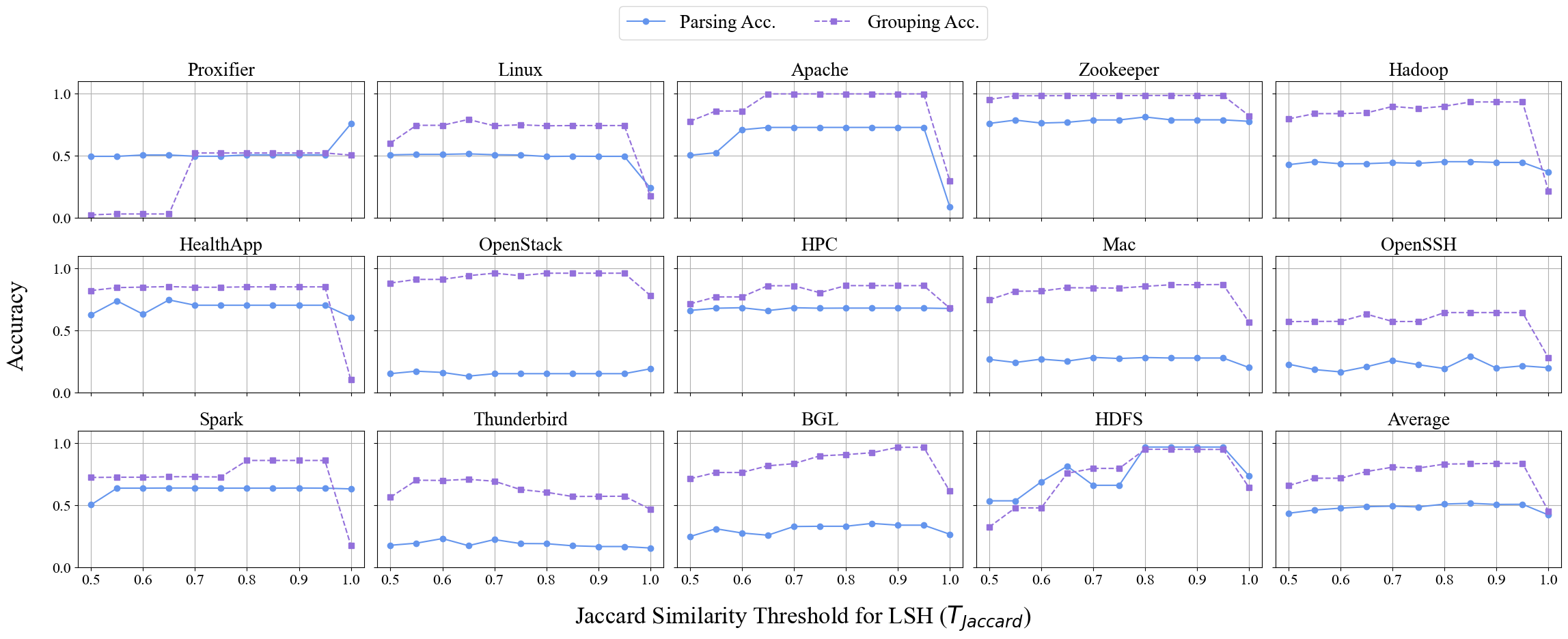}
    \caption{Impact of LSH Jaccard Threshold on PA and GA.}
    \Description{}
    \label{fig:comparison of different jaccard threshold}
\end{figure*}

As shown in Figure~\ref{fig:comparison of different jaccard threshold}, while the optimal Jaccard threshold varies across datasets due to their distinct characteristics, the parsing performance is relatively insensitive to a considerable range of the Jaccard threshold (e.g., 0.8 - 0.95). In particular, \textbf{setting $T_{\text{Jaccard}} = 0.9$, we can achieve near-optimal parsing performance across all log datasets}; the exception is the Thunderbird log, maybe because its log messages are relatively short thus the Jaccard similarities between logs are smaller when there are variables).  For most datasets, performance drops sharply when ($T_{\text{Jaccard}}=1$), highlighting the influence of variable parts that persist after filtering with the regex pattern, \texttt{\^{}[a-zA-Z]+[.,]*\$}. %The proportion of these variable parts within the shingles determines the optimal threshold for each dataset. 

GA is more sensitive to the Jaccard threshold because it relies on precise log similarity evaluation for clustering. A strict threshold ($T_{\text{Jaccard}}=1$) risks under-grouping by ignoring minor variations, while a relaxed threshold ($T_{\text{Jaccard}}<1$) may cause over-grouping by merging dissimilar logs. Both extremes negatively affect GA, emphasizing the need for careful threshold tuning. In contrast, PA focuses on identifying static and dynamic components within logs and is less affected by similarity evaluation. %This distinction underscores the importance of balancing GA and PA through dataset-specific parameter optimization.

\begin{tcolorbox}[colframe=blue!60!black, colback=blue!5!white, coltitle=black, sharp corners=southwest, width=\linewidth, boxrule=0.2mm, enlarge left by=0mm, enlarge right by=0mm, boxsep=0pt]
%\textbf{Answer to RQ1:}\\
LogLSHD outperforms baseline parsers in accuracy and efficiency. 
Our sensitivity analysis of the Jaccard similarity threshold ($T_{\text{Jaccard}}$) indicates that a fixed setting (e.g., $T_{\text{Jaccard}} = 0.9$) can achieve near-optimal performance across all log datasets in the Loghub-2.0 benchmark.
%It faces challenges on Proxifier and Linux datasets due to template similarity and complexity. The optimal similarity threshold varies by dataset, impacting accuracy and grouping.
\end{tcolorbox}

\subsection{RQ2: What is the impact of different grouping strategies on the performance of LogLSHD?} 
\label{sec:rq2}

\subsubsection{Motivation}
%\xingfang{please check, not sure if it is correct.}
The study aims to evaluate the impact of different grouping strategies on the performance of LogLSHD. In particular, it investigates how initially grouping %\xingfang{grouping is better here? if the RQ only considers the five-character grouping, should specific in the RQ rather than utilizing the general ``grouping strategies"}\shuwei{revised a little, not sure whether ``five-character grouping" is better here?} 
logs using token count, log content length, and specific-position characters (as described in Section~\ref{sec:grouping_strategies}) affects Parsing Accuracy (PA), Grouping Accuracy (GA), and parsing time. The research question seeks to understand the trade-offs between grouping precision and computational efficiency, especially for datasets with varying levels of complexity and variability.

\subsubsection{Approach}
As mentioned in Section~\ref{sec:grouping_strategies}, the experiments utilized token count and log content length of the log as the base grouping criteria. Additionally, different combinations of specific-position character matching between log messages are evaluated: the first, 25\%, 50\%, 75\%, and last characters.
%fixed five-character segments were extracted from specific positions within each log (the first, 25\%, 50\%, 75\%, and last characters) to examine the impact of grouping strategies.

\subsubsection{Result}

\begin{figure*}
    \centering
    \includegraphics[width=1\linewidth]{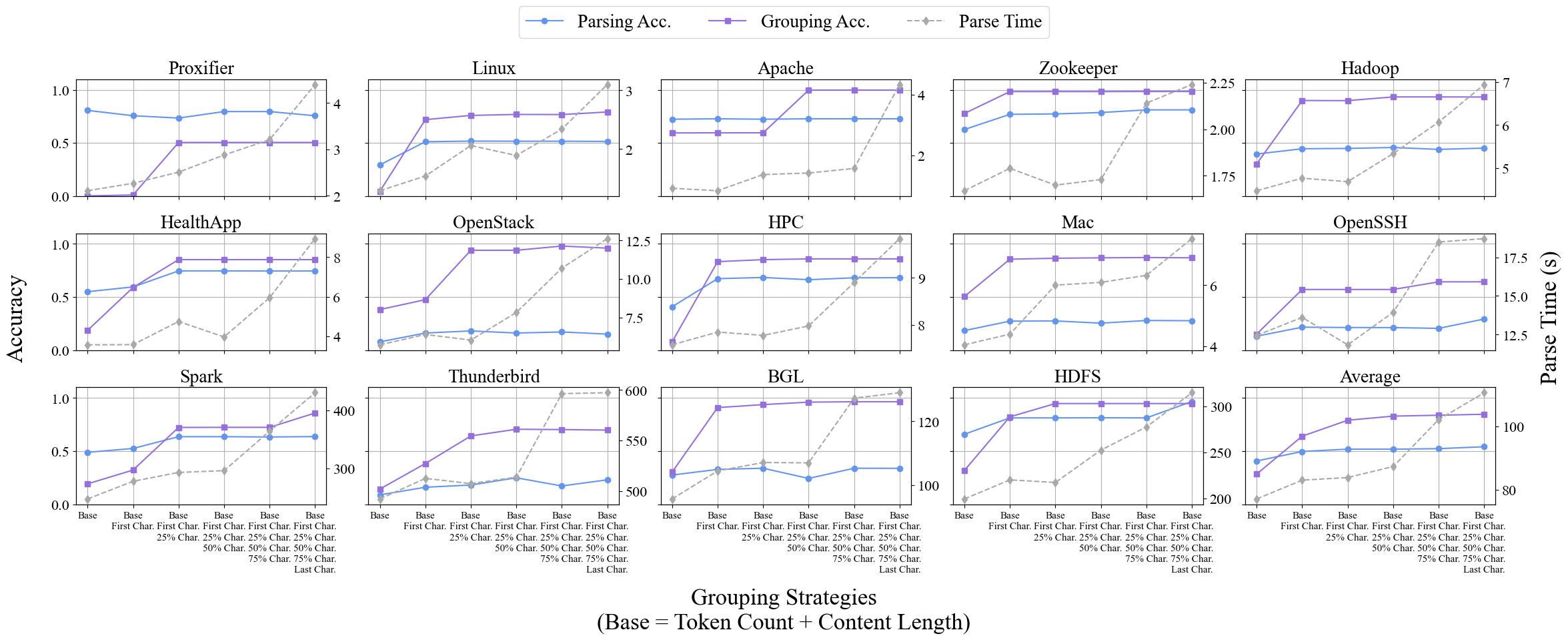}
    \caption{Impact of Grouping strategies on PA and GA.}
    \Description{}
    \label{fig:comparison_grouping_strategies}
\end{figure*}

Figure~\ref{fig:comparison_grouping_strategies} presents the effects of different grouping strategies on PA, GA, and parsing time across various datasets, using the optimal $T_{\text{Jaccard}}$ as identified in Table~\ref{tab:LogLSHD_acc}.

PA generally exhibited increasing trends across datasets when applying finer grouping strategies. Datasets such as Proxifier, Apache, and Zookeeper showed minimal changes in PA, reflecting their low variability, where simple grouping strategies were sufficient to capture the main patterns. In contrast, datasets with a larger number of unique templates or annotated logs, such as HealthApp and Thunderbird, exhibited more significant fluctuations in PA, highlighting the impact of grouping strategies on these datasets.

GA generally improved with more refined grouping strategies across most datasets, suggesting that finer grouping effectively captured similarities among logs. Parsing time, however, increased with the complexity of grouping strategies. Smaller datasets, such as Proxifier and Linux, maintained relatively stable parsing times due to their lower computational demands. In contrast, larger datasets, such as Thunderbird and Spark, experienced significant increases in parsing time, particularly under more detailed grouping strategies.

On average, initial grouping strategies resulted in lower GA, but substantial improvements were observed with the addition of character pick-up in grouping strategies. In this experiment, the Base strategy combined with first character, 25\% character, and 50\% characters achieved near-optimal performance. Parsing time exhibited an overall linear growth trend, though potential bottlenecks emerged when strict grouping strategies were applied to larger datasets.

\begin{tcolorbox}[colframe=blue!60!black, colback=blue!5!white, coltitle=black, sharp corners=southwest, width=\linewidth, boxrule=0.2mm, enlarge left by=0mm, enlarge right by=0mm, boxsep=0pt]
%\textbf{Answer to RQ2:}\\
Finer initial grouping improved PA and GA, especially for complex datasets, but increased parsing time, particularly for larger datasets. The Base strategy with the matching of the first, 25\%, and 50\% characters achieved near-optimal performance.
\end{tcolorbox}

\subsection{RQ3: What is the impact of the DTW-based template extraction method in LogLSHD?}
\label{sec:rq3}

\subsubsection{Motivation}
This research question aims to examine the impact of incorporating Dynamic Time Warping (DTW) in the template extraction step of LogLSHD. % to improve template generation accuracy. 
The goal is to assess the trade-off between accuracy and parsing time when using DTW in the template extraction.

\subsubsection{Approach}
To evaluate the impact of incorporating Dynamic Time Warping (DTW) into the template extraction step, we compare the performance of LogLSHD with DTW to LogLSH, the variant of LogLSHD that does not use DTW and relies on standard token composition for template extraction. Both methods are tested across multiple datasets, focusing on parsing accuracy (PA) and parsing time. The Jaccard thresholds are configured as shown in Table~\ref{tab:LogLSHD_acc}. %This comparison allows us to assess the effect of DTW on template generation accuracy and parsing efficiency, examining the trade-off between improved accuracy and the increased computational cost introduced by DTW.

%As mentioned in Section~\ref{sec:Hyperparameters of LSH}, we examine a range of Jaccard thresholds of LogLSHD to find the optimal one based on the highest $PA+GA$ for each dataset, as the result is shown in Table~\ref{tab:LogLSHD_acc}. 

\subsubsection{Result}

Table~\ref{tab:loglsh_vs_loglshd} compares the performance of LogLSH and LogLSHD. From the results, we find that incorporating DTW in the template extraction step leads to a 22.7\% improvement in PA, demonstrating its effectiveness in enhancing template generation accuracy. However, this improvement comes at the cost of a 9.47\% increase in parsing time, attributed to the additional computational overhead introduced by DTW. Meanwhile, GA shows no significant enhancement, as DTW does not directly influence the grouping process. These results suggest that the trade-off between parsing time and accuracy is favorable, particularly for applications where precision in template generation is critical.

\begin{table}[ht]
    \centering
    \caption{Performance comparison between LogLSH and LogLSHD.}
    \label{tab:loglsh_vs_loglshd}
    \begin{tabular}{@{}lccc@{}}
        \toprule
        \textbf{Metric(Avg.)} & \textbf{LogLSH} & \textbf{LogLSHD} & \textbf{Relative Change (\%)} \\
        \midrule
        PA & 0.44 & 0.54 & +22.7 \\
        GA & 0.83 & 0.85 & +2.4 \\
        Parsing Time(s) & 99.56 & 109 & +9.47 \\
        \bottomrule
    \end{tabular}
\end{table}

\begin{tcolorbox}[colframe=blue!60!black, colback=blue!5!white, coltitle=black, sharp corners=southwest, width=\linewidth, boxrule=0.2mm, enlarge left by=0mm, enlarge right by=0mm, boxsep=0pt]
%\textbf{Answer to RQ3:}\\
Incorporating DTW into LogLSHD improves the average parsing accuracy by 22.7\%, but at the cost of a 9.47\% increase in parsing time, with no significant impact on grouping accuracy.
\end{tcolorbox}
\section{Threats to Validity}
\label{sec:threats_to_validity}

This section discusses the potential limitations of LogLSHD, focusing on the challenges of manual parameter tuning and sensitivity to noise during template extraction, which impact its adaptability and generalizability across diverse datasets.

\noindent \textbf{LSH Parameter Configuration.}
Despite the efficiency advantages of LogLSHD, the performance of LSH relies on parameter configuration for different datasets. Specifically, adjusting the Jaccard similarity threshold is important for achieving optimal results. %Moreover, the proportion of variable parts within logs varies significantly not only across datasets but also within the same dataset. This variability complicates the application of a fixed Jaccard threshold, as it fails to adapt to the diverse characteristics of log data. 
Nevertheless, our experiments show that we can still find a fixed parameter ($T_{\text{Jaccard}}=0.9$) that achieves near optimal parsing performance across all studied datastes.
Designing automated or adaptive strategies for LSH parameter optimization could benefit our approach.

\noindent \textbf{Sensitivity to Noise in Log Template Extraction.}
A potential threat to the validity of our approach lies in its sensitivity to noise during template extraction. Specifically, when log templates from different categories are mistakenly grouped together, they may introduce noise into the template generation process. This issue can lead to the propagation of errors, where misclassified logs result in entirely incorrect templates. %Although efforts were made to minimize such noise, our method's ability to generalize across diverse datasets remains limited. 
Future work could develop more robust mechanisms to handle noise and improve the robustness of template extraction.
\section{Related Work}
\label{sec:related_work}

This section reviews log parsing methods, categorizing them into statistic-based and learning-based approaches.

\subsection{Statistic-Based Log Parsers}

Statistic-based log parsers generate templates based on statistical characteristics of log messages, such as token frequency and similarity measures~\cite{vaarandi2003data, vaarandi2015logcluster, hamooni2016logmine}. These methods are typically categorized into frequency-based, similarity-based, and heuristic-based approaches~\cite{jiang2024large}. Frequency-based parsers analyze the occurrence frequency of tokens to distinguish static and dynamic parts of log messages, offering a straightforward approach to template generation~\cite{dai2020logram, dai2023pilar, vaarandi2003data, nagappan2010abstracting}. Similarity-based parsers employ clustering algorithms to group log messages based on predefined distance metrics, summarizing templates for each cluster~\cite{fu2009execution, mizutani2013incremental, tang2011logsig}. Heuristic-based parsers utilize rule-based strategies or specialized data structures, such as parsing trees or longest common subsequence algorithms, to identify patterns and generate templates~\cite{du2016spell, he2017drain, makanju2009clustering}. Evaluations of these parsers indicate that while they achieve high grouping accuracy, their parsing accuracy can be limited, particularly when handling highly dynamic or unstructured logs~\cite{jiang2024large, yu2023brain, messaoudi2018search}.

% Statistic-based log parsers form templates according to the statistical facts in log messages, such as token frequencies~\cite{jiang2024large}. These parsers can be further categorized as frequency-based, similarity-based, or heuristic-based. Frequency-based parsers such as Logram, distinguish the static and dynamic parts in log messages with their existence frequency~\cite{dai2020logram, vaarandi2003data, nagappan2010abstracting, vaarandi2015logcluster}. Based on different definitions of distances and similarity, similarity-based parsers group log messages with clustering algorithms, and then summarize a template for each group~\cite{fu2009execution, hamooni2016logmine, mizutani2013incremental, shima2016length, tang2011logsig}. Heuristic-based parsers group the log messages and summarize templates based on different heuristic algorithms or data structures such as parsing trees or longest common subsequence-based approach~\cite{du2016spell, he2017drain, makanju2009clustering, messaoudi2018search, yu2023brain}. In the large-scale evaluation by Jiang \textit{et al.}~\cite{jiang2024large}, these parsers share a similar preprocessing stage but vary in parsing algorithms. According to the reported results, statistic-based log parsers can reach high grouping accuracies but are generally low in parsing accuracies.

\subsection{Learning-Based Log Parsers}

Learning-based log parsers employ machine learning or deep learning techniques to extract parsing patterns from complex or evolving log formats adaptively. Sequence-based models, such as LSTMs, are commonly used to handle unseen log formats by embedding log messages and inferring event templates~\cite{rucker2022flexparser, wu2024logptr}. %Advanced approaches, like FlexParser, utilize stateful LSTM to capture evolving patterns and extract templates over training epochs~\cite{rucker2022flexparser}. 
Other solutions adopt transfer learning and self-supervised learning, such as using masked-language modeling to dynamically generate templates for new logs~\cite{liu2022uniparser,le2023log}. Recent advancements also incorporate Natural Language Processing (NLP) in Semantic-based Log Parsers to enhance semantic understanding, further improving performance on unstructured logs~\cite{aussel2018improving, liu2022uniparser, le2023log, li2023did, ma2024llmparser, pei2024self}. While these methods offer high adaptability, they require significant annotated data and computational resources, posing challenges in practical deployment. Thus, ongoing research continues to explore lightweight architectures and unsupervised techniques to address these limitations.
\section{Conclusion}
\label{sec:conclusion}

This study addresses the challenges of scalability and accuracy in log parsing by proposing an efficient method, LogLSHD. Building upon the strengths of static-based methods, LogLSHD leverages the LSH algorithm with a grouping strategy to manage large-scale log data efficiently and incorporates DTW to enhance template extraction accuracy. Experimental results demonstrate that LogLSHD achieves superior parsing accuracy, representing a notable 22\% average improvement over AEL and a 15\% average improvement over Drain. Additionally, LogLSHD excels in efficiency, achieving a significant 93\% reduction in average parsing time compared to AEL and a 73\% reduction compared to Drain.
%However, the performance of LogLSHD heavily depends on manual parameter tuning, particularly the Jaccard similarity threshold in LSH. The varying proportion of variable parts both across and within datasets complicates the use of a fixed Jaccard similarity threshold, thereby limiting its adaptability. Additionally, the method's sensitivity to noise during feature extraction constrains its generalizability across diverse datasets. 
Future research could develop automated or adaptive parameter optimization strategies %and robust noise-handling mechanisms 
to enhance the performance and robustness of LogLSHD.

% %%
% %% The next two lines define the bibliography style to be used, and
% %% the bibliography file.
\balance
\bibliographystyle{ACM-Reference-Format}
\bibliography{main}

\end{document}